
%
%
%
\input harvmac.tex


\lref\ka {For a review see I.R. Klebanov and A. Pasquinucci, lectures given at
Trieste Summer School of Theoretical Physics, 1992, hepth/9210105.}
\lref\rev {For a review see I.R. Klebanov, in {\it String Theory and
Quantum Gravity '91} World Scientific 1992,
and references therein.}
\lref\lee {J.C. Lee Phys. Lett. {\bf B337} (1994) 69;
Phys. Lett. {\bf B326} (1994) 79;
Prog. Th. Phys. Vol. {\bf 91.2} (1994) 353.}
\lref\post{A.M. Polyakov, Mod. Phys. Lett. {\bf A6} (1991) 635.}
\lref\kp {I.R. Klebanov and A.M. Polyakov, Mod. Phys. Lett. {\bf A6} (1991)
3273.}
\lref\lz{B. Lian and G. Zuckman, Phys. Lett {\bf B266} (1991) 21.}
\lref\wit{E. Witten, Nucl. Phys. {\bf B373} (1992) 187;
E. Witten and B. Zweibach, Nucl. Phys. {\bf B377} (1992) 55.}
\lref\aj{J. Avan and A. Jevicki, Phys. Lett. {\bf B266} (1991) 35,
{\bf B272} (1991) 17.}
\lref\poly{A.M. Polyakov, Princeton preprint PUPT-1289,
Lectures given at 1991 Jerusalem Winter School,
{\it Jerusalem Gravity 1990}, 175.}
\lref\gkn{D.J. Gross, I.R. Klebanov and M.J. Newman, Nucl. Phys.
{\bf B350} (1991) 621;
D.J. Gross and I.R. Klebanov, Nucl. Phys. {\bf B352} (1991) 671;
K. Demetrifi, A. Jevicki and J.P. Rodrigues, Nucl. Phys. {\bf B365} (1991)
499 ;
U.H. Danielson and D.J. Gross, Nucl. Phys. {\bf B366} (1991) 3. }
\lref\gold{Goldstone, unpublished; V.G. Kac, in {\it Group Theoretical
Methods in Physics, Vol 94}, Springer-Verlag, (1979);
G. Segal, Comm. Math. Phys. {\bf 80} (1981) 301.}
\lref\eky{T. Eguchi, H. Kanno and S. Yang, Phys. Lett {\bf B298}
(1993) 73.}
\lref\dhk{J. Distler, Z. Hlousek and H. Kawaii, Int.J.Mod.Phy. Vol {\bf 5.2}
(1990) 391.}
\lref\ohta{ K. Itoh and N. Ohta, Nucl. Phys. {\bf B377} (1992) 113,
Prog.Th.Ph.Sup. {\bf 110} (1992) 97.}
\lref\bmp{P. Bouwknegt, J. McCathy and K. Pilch, Nucl. Phys. {\bf B377} (1992)
514.}
\lref\cl{T. Chung and J.C. Lee, NHCU-HEP-94-33, hepth/9412095, to appear in
Phys. Lett. B.}
\lref\ward{I. Klebanov, Mod. Phys. Lett. {\bf A7} (1992) 723,
S. Kachru, Mod.Phys.Lett. {\bf A7} (1992) 1419.}
\lref\pari{E. Marinari and G. Parisi, Phys. Lett. {\bf B240} (1990) 375.}

\font\blackboard=msbm10 \font\blackboards=msbm7
\font\blackboardss=msbm5
\newfam\black
\textfont\black=\blackboard
\scriptfont\black=\blackboards
\scriptscriptfont\black=\blackboardss
\def\blackb#1{{\fam\black\relax#1}}


\def\del {{\partial}}
\def\half {{1 \over 2 }}
\def\ket {{\rangle}}
\def\ZZ {{ \blackb Z }}
\def\bz {{ \bf z}}

\def\boo{{ \bf 0}}
\def\bx {{ \bf X}}
\def\bfi {{ \bf \Phi}}
\def\bd {{\bf D}}
\def\bps {{\bf \Psi}}
\def\bs {{\bf S}}
\def\bdel {{\bf \Delta}}
\def\bg {{\bf G}}


\Title{\vbox{\baselineskip12pt\hbox{NHCU-HEP-95-06}
    \hbox{May 1995}
    \hbox{hep-th/9505107}
    }}
{\vbox{\centerline{Superfield Form of Discrete Gauge States}
	\vskip 2mm\centerline{in $\hat c=1$ 2d Supergravity}}}

\vskip .5cm
\centerline{Tze-Dan Chung\footnote{$^\dagger$}
{\it email: chung@math.nctu.edu.tw} and Jen-Chi Lee\footnote{$^*$}
{\it email: jclee@twnctu01.bitnet}}
\bigskip\centerline{\it Department of Electrophysics}
\centerline{\it National Chiao Tung University}
\centerline{\it Hsinchu, Taiwan 30050, R.O.C.}

\vskip 2cm
\rm
\noindent

\centerline{\bf Abstract}

A general formula for the discrete states (Neveu-Schwarz sector) in $N=1$
$2D$ super-Liouville theory is written down in
the world-sheet supersymmetric form.
We then derive a set of gauge states at the discrete momenta.
These discrete gauge states are shown to carry the $w_\infty$ charges and
serve as the symmetry parameters in the old covariant quantization
of the theory.

\Date {}


\newsec {Introduction}

One of the main motivations to study 2d quantum gravity was to
understand the non-perturbative information of string theory.
The discretized matrix model \refs{\rev} approach developed so far has been
very sucessful. On the other hand, the continuum Liouville
theory serves \refs{\ka} as an important consistency check for the matrix model
approach. Since little is now
known for supersymmetric matrix model \refs{\pari}, it would be
interesting to develop 2D super-Liouville theory \refs{\bmp}\refs{\ohta}
and compare its results
directly with the high dimensional critical string theory.

In the Liouville theory, in addition to the massless tachyon mode, an
infinite number of massive discrete states were discovered
\refs{\gkn}\refs{\post} and the target
space-time $w_\infty$ symmetry \refs{\wit}\refs{\kp}\refs{\aj}
and Ward identities \refs{\ward} were then identified.
In a previous paper \refs{\cl},
we introduced the concept of discrete gauge states (DGS) and gave a general
formula for them. These DGS were then shown to carry the
$w_\infty$ charges and can be
considered as the symmetry parameters in the old covariant
quantization of the theory. This is in parallel with the
BRST approach \refs{\bmp}\refs{\wit}\refs{\ohta}\refs{\lz}
appeared in the literature,
and can be compared with the works of 10D critical string theory \refs{\lee},
where a complete gauge state analysis turns out to be extremely difficult
to attain.
In this paper, we will generalize our results in \refs{\cl}
to $N=1$ super-Liouville theory in
the worldsheet supersymmetric way.
We will work out the DGS of the Neveu-Schwarz sector in the zero ghost
picture.
We organize the paper as following.
In section II, we discuss the $N=1$
super-Liouville theory and set up the notations.
In section III, we calculate the general formula for discrete states in a
worldsheet superfield form which seems missing in the liturature. The DGS and
$w_\infty$ charges were then given in section IV.
A brief conclusion was summarized in the final section.

\newsec{2D Super-Liouville Theory}

The $N=1$ two dimensional supersymmetric Liouville action is
given by \refs{\dhk}
\eqn\act{
S={1\over 8 \pi} \int d^2 \bz [ g^{\alpha\beta}
(\del_\alpha \bx \del_\beta \bx + \del_\alpha \bfi \del_\beta \bfi)
-Q {\hat {\bf Y}} \bfi ], }
where $\bfi$ is the super-Liouville field, $\hat {\bf Y}$ the
superfield curvature,
$d\bz =dzd\theta$
and with $\bx ^\mu= \pmatrix{\bfi \cr \bx \cr}$,
\eqn\sdef{{ \bx ^{\mu}(z,\theta,\bar z,\bar\theta)
 = X^{\mu} + \theta \psi ^{\mu} +
\bar\theta \bar\psi ^{\mu}
+\theta\bar\theta F^ \mu }.}
Bold faced variables denote superfields hereafter.

For $\hat c=1 = {2\over 3} c$ theory $Q$,
which represents the background charge of the super-Liouville field,
is set to be $2$ so that the total conformal anomaly
cancels that from conformal and superconformal ghost contribution.

The equations of motion show that the left and right-moving
components of $\bx^{\mu}$ decouple,
and the auxiliary fields $F ^\mu$ vanish. As a result,
we need to consider only one of the chiral sectors, while
the other (anti-holomorphic) sector has a similar formula.
The stress energy tensor is
\eqn\seten{{\bf T}_{zz} = - \half \bd \bx^ \mu \bd ^2 \bx_{\mu}
- \half Q \bd^3 \bfi =T_F +\theta T_B,}
with
\eqn\tf{\eqalign{T_F=& -\half \del X^{\mu} \del X_{\mu} -\half Q \del^2 X^0
+\half \psi^{\mu} \del  \psi_{\mu} \cr
T_B=& -\half \psi^{\mu} \del X_ {\mu}- \half Q\del \psi^0  ,}}
where  $\bd=\del_\theta + \theta \del_z$, and now
$\bx^\mu = X^\mu (z) + \theta \psi^\mu (z)$.

For the Neveu-Schwarz sector, if we define the mode expansion by
\eqn\xpmod{
\del_z  X^\mu = - \sum_{n=-\infty}^{\infty}z^{-n-1}
(\alpha_n^0 , i \alpha_n^1 ),}
\eqn\psmod{
\psi^\mu = - \sum_{r \in {\ZZ +\half}} z^{-r-\half}
(b_r^0 , i b_r^1 ),}
then we have
\eqn\modex{ [ \alpha^{\mu}_m, \alpha^\nu_n ] = n \eta^{\mu\nu}\delta_{m+n},
\qquad \{b^\mu_r, b^\nu_s \} =  \eta^{\mu\nu}\delta_{r+s}.}
With the Minkowski metric $\eta_{\mu\nu}= \pmatrix{-1& 0 \cr 0 &1 \cr}$,
$Q^\mu=\pmatrix{2 \cr 0 \cr}$ and the zero modes
$\alpha^{\mu}_0 = f^\mu = \pmatrix {\epsilon \cr p \cr}$, we find the
super-Virasoro generators as modes of $T_F$ and $T_B$,
\eqn\vir{\eqalign{
L_n =& ( {n+1 \over 2} Q^\mu + f^\mu) \alpha_{\mu,n}
+ \half \sum_{k \neq 0} : \alpha_{\mu,-k} \alpha_{n+k}^\mu :
+\half \sum_{r \in {\ZZ +\half}}(r+n+\half):b_{-r}^\mu b_{n+r,\mu}: \cr
L_0 =& \half (Q^\mu + f^\mu) f_\mu + \sum_{k=1}^{\infty}
: \alpha_{\mu,-k} \alpha_k^\mu :
+\half \sum_{r \in {\ZZ +\half}}(r+\half):b_{-r}^\mu b_{r,\mu}:
, \cr
G_r=&\sum_{s \in {\ZZ +\half}}  \alpha^\mu_{r-s}b_{\mu,s}
+(r+\half)Q^\mu b_{\mu,r}}.}

The vacuum $|0 \ket$ is annihilated by all $\alpha_n^\mu$ and $b_r^\mu$
with $n>0$ and $r>0$. In the old covariant quantization of the theory,
physical states $|\psi \ket $ are those satisfying the conditions
\eqn\phyc{\eqalign{
G_\half |\psi \ket &=
G_{3\over 2} |\psi \ket = 0 \cr
and \qquad L_0 |\psi\ket  &= \half |\psi\ket .}}

\newsec {World-sheet Superfield Form of The Discrete States}

With \seten\  one can easily check that the two branches of
massless ``tachyon''
\eqn\tahcbr{T^{\pm}(p) = \int d\bz e^{ip\bx +(\pm |p| -1) \bfi} }
are positive norm physical states.
It was also known that there exists discrete momentum physical states.
Writing $\int d \bz \bps_{J,\pm J}^{(\pm)}= T^{(\pm)}({ \pm J})$,
the discrete states in the ``material gauge'' are
\eqn\disc{\bps_{J,M}^{(\pm)} \sim (H^-)^{J - M} \bps_{J,J}^{(\pm)}
\sim (H^+)^{J + M} \bps_{J, -J}^{(\pm)} .}
where
\eqn\sutwo{H^{\pm}={\sqrt 2}\int d\bz e^{\pm i \bx(\bz)},
\qquad H^0 = \int d\bz \bd \bx .}
are zero modes of the level $2$ $SU(2)_{\kappa =2}$
Kac-Moody algebra in
$\hat c=1$ 2d superconformal field theory.
Here we note that the NS sector corresponds to states with $J\in \ZZ$
while the Ramond sector corresponds to those with $J\in \ZZ+\half$.

To find the explicit expressions for the discrete states,
we first define the super-Schur polynomials,
\eqn\schdef{
\bs _k \left( -i\bx \right) = {\bd^k e^{-i\bx} \over [k/2]!} e^{i\bx},}
where $[{k\over 2}]$ denotes the integral part of ${k\over 2}$,
as the N=1 generalization to the Schur polynomial $S_k$,
which is defined as
\eqn\schur{
Exp \left ({\sum_{k=1}^{\infty} a_k x^k}\right)
= \sum_{k=0}^{\infty} S_k \left( \{a_k\} \right) x^k .}
Note that $S_k (\{-i \del^m \bx / m!\})=\bs_{2k}(-i\bx)$.
Direct integration shows that
\eqn\lemone{\eqalign{
\int d\bz_1 {1\over (z_1 -z -\theta_1\theta)^n}f(\bx_1)
=&{\bd^{2n-1}f(\bx) \over (n-1)!} \cr
=& {\del^{n-1}_z (f'(X)\psi) + \theta \del^n_z f(X)\over (n-1)!}} }
Using \lemone\ we obtain
\eqn\psia{\eqalign{
\bps_{J,J-1}^{\pm} &\sim \bs_{2J-1}(-i\bx) e^{i(J-1)\bx + (\pm J-1)\bfi} \cr
&= {1\over (J-1)!} [-i \del^{J-1}(e^{-iX^1} \psi^1)
+ \theta \del^{J-1}e^{-iX^1}]
e^{iJ\bx +(\pm J-1)\bfi}. }}
For example, by
\eqn\corres{(-\bd^{2r} \bfi^0 ,
i\bd^{2r} \bx^1 )
\to b_{-r}^{\mu},
\qquad (-\bd^{2n} \bx^0,
i\bd^{2n}\bx^1) \to \alpha^{\mu}_{-n}}
we have
\eqn\exone{\bps_{1,0}^+=\bd\bx \to b_{-\half}^1 | f^{\mu}=(0,0)\ket}
and
\eqn\extwo{\eqalign{
\bps_{2,\pm 1}^+ =&[-i\bd^3\bx-\bd\bx \bd^2\bx]e^{\pm i \bx + \bfi} \cr
\to &[- b_{-{3\over 2}}^1 +  b_{-\half}^1 \alpha_{-1}^1]
 |f^{\mu}=(1,\pm 1) \ket .}}
They can be checked to satisfy the physical state conditions in \phyc .

Performing the operator products in \disc\ ,
the discrete states $\bps_{J,M}^\pm$ are
\eqn\psieva {\eqalign{
\bps_{J,M}^\pm \sim & \prod_{i=1}^{J-M} \int {d\bz_i } \bz_{i0}^{-J}
\prod_{j<k}^{J-M} \bz_{jk} \cr
& Exp \left[ \sum_{i=1}^{J-M}\big[ -i  \bx(\bz_i)\big] +
 \big( iJ\bx(\bz_0) +
(-1 \pm J) \bfi (\bz_0)\big)\right],}}
where $\bz_{ab}=z_a-z_b-\theta_a\theta_b$.
If we write $\bz_{ab}=\bz_{a0}-\bz_{b0}-
(\theta_a-\theta_0)(\theta_b-\theta_0)$,
and use $\int {d\bz_a (\theta_a-\theta_0) \bz_{a0}^{-n}}
f(\bx_a)= \bd^{2n-2}f(\bx_0)/(n-1)!$, we get, for $M=J-2$,
\eqn\psimt{
\bps_{J,J-2}^{\pm} \sim [2\bs_{2J-3}\bs_{2J-1} + \bs_{2J-2}\bs_{2J-2}]
 e^{i(J-2)\bx + (\pm J-1)\bfi}.}
The vertex operators correspond to the upper components of \psimt , i.e.,
\eqn\psibb{\eqalign{
\int d\theta \bps_{J,J-2}^{\pm} \sim &
[ \left( iJ\psi^1 +(\pm J-1)\psi^0 \right)
(S^2_{J-1} + 2 S_{J-{3\over2}}^{NS} S_{J-\half}^{NS}) \cr
&- 2 J(S_J S_{J-{3\over2}}^{NS} - S_{J-1} S_{J-\half}^{NS})]
e^{i(J-2)X^1+(\pm J -1) X^0},}}
where  $S_J=S_J (\{-i \del^m \bx / m!\})$ and
$S^{NS}_{k+\half}= \sum_{m=0}^k {-i S_m \del^{k-m}\psi^1 \over (k-m)!}$.
Using \corres \ and \psimt\ it is found that
\eqn\pstwon{\bps_{2,0}^+ \to
[2b_{-\half}^1 b_{-{3\over 2}}^1
+\alpha_{-1}^1 \alpha_{-1}^1] |f^{\mu}=(1,0) \ket .}
It can be checked that it satisfies the physical state conditions \phyc .

 For $M=J-3$, a straighforward calculation gives
\eqn\psib{\eqalign{
\bps_{J,J-3}^{\pm} \sim &[3!\bs_{2J-1}\bs_{2J-3}\bs_{2J-5}
+ 3!\bs_{2J-2}\bs_{2J-3}\bs_{2J-4} \cr
& - {3! \over 1!2!} \bs_{2J-1}\bs_{2J-4}^2
- {3! \over 2!1!} \bs_{2J-2}^2\bs_{2J-5}]
e^{i(J-3)\bx + (\pm J-1)\bfi}. }}

It is now easy to write down an expression for general $M$,
\eqn\allpsi{
\bps_{J,M}^\pm  \sim
\left|\matrix {
\bs_{2J-1} & \bs_{2J-2}   & \cdots & \bs_{J+M} \cr
\bs_{2J-2} & \bs_{2J-3}   & \cdots & \bs_{J+M-1} \cr
\vdots   & \vdots    & \ddots & \vdots \cr
\bs_{J+M}  & \bs_{J+M-1}  & \cdots & \bs_{2M+1} \cr
} \right| '
Exp \left[ \big( iM \bx(\bz_0) +
(-1 \pm J) \bfi(\bz_0)\big)\right],}
with $\bs_k = \bs_k \left(  -i \bx(\bz_0) \right)$ and
$\bs_k = 0$ if $k<0$.
We will denote the rank $(J-M)$
``primed''-determinant in \allpsi\ as
$\bdel'(J,M,-i \bx)$,
which (by definition) has all the signed terms in the normal determinant,
except with a multiplicity of the multinomial coefficient
$(J-M)!\over n_a!n_b!...$ for the term $\bs^{n_a}_a\bs^{n_b}_b ...$
(where $\sum_a n_a=J-M$).

\newsec{DGS and $w_\infty$ Charges}

It was known \refs{\bmp}\refs{\ohta} that the discrete states in \disc\
satisfy the $w_\infty$ algebra
\eqn\winftp{ \int d\bz
\bps_{J_1,M_1}^+(\bz) \bps_{J_2,M_2}^+(\boo) = (J_2 M_1 - J_1 M_2)
\bps_{J_1+J_2-1,M_1+M_2}^+(\boo),}
\eqn\winftm{
\int d \bz
\bps_{J_1,M_1}^-(\bz) \bps_{J_2,M_2}^-(\boo) \sim 0,}
where the RHS is defined up to a DGS.

In general, there are two types of gauge states in the old covariant
quantization of the theory,

Type I:
\eqn\gauone{
|\psi \ket = G_{-\half} |\chi \ket \qquad where \qquad G_{\half}|\chi \ket
= G_{3\over 2}|\chi \ket =L_0 |\tilde\chi \ket = 0 }

Type II:
\eqn\gautwo{\eqalign{
|\psi \ket =\left(G_{-{3\over 2}} + 2L_{-1}G_{-\half}\right)|\tilde\chi \ket
\qquad where \qquad &G_{\half} |\tilde\chi \ket =
G_{3\over 2} |\tilde\chi \ket =0 \cr
& (L_0+1) |\tilde\chi \ket = 0 .}}
They satisfy the physical state conditions \phyc , and have zero norm.
There is an infinite number of continuum momentum gauge state solutions
for \gauone\ and \gautwo\ . However, as far as the dynamics is concerned,
we are only interested in those with discrete momentum.

At mass level one, $ f_\mu (f^\mu +Q^\mu) = 0$,
only gauge states of type I are found:
$f_\mu \alpha^\mu_{-1} |f \ket$,
where $ |f \ket = :e^{ip \bx + \epsilon \bfi} :|0 \ket$.
The DGS $G^-_{1,0} = :\bd \bfi e^{-2 \bfi}:|0 \ket $
corresponds to the momentum of $\bps^-_{1,0}$. There is no corresponding
DGS for $\bps^+_{1,0}=:\bd \bx :$.

At the next mass level,
$ f_\mu (f^\mu +Q^\mu) = -2$, $N_{\mu\nu}=-N_{\nu\mu}$ and
$M_\mu=2N_{\mu\nu}(f^\nu+Q^\nu)$,
the type I gauge state is found to be
\eqn\ogauo{
|\psi\ket = [(M_\mu f_\nu \alpha_{-1}^\mu b_{-\half}^\nu
+ M_\mu b_{-{3\over 2}}^\mu +
2N_{\mu\nu}\alpha_{-1}^{\mu}b_{-\half}^\nu]|f\ket ,}
while the type II state is
\eqn\ogaut{|\psi\ket = [(2 f_\mu f_\nu +  \eta_{\mu\nu} )
\alpha_{-1}^\mu b_{-\half}^\nu +
(3f_\mu - Q_\mu )b_{-{3\over 2}}^\mu] |f\ket .}

As in the bosonic Liouville theory \refs{\cl},
the gauge states corresponding to the discrete momenta of
$\bps^+_{2,\pm 1}$ are degenerate, i.e.,
the type I and type II gauge states are linearly dependent:
\eqn\dgsaa{G^+_{2,\pm 1} \sim \left[
\pmatrix{ 1& \mp 2 \cr
	  \mp 2 & 3 \cr }
\alpha_{-1}^\mu b_{-\half}^\nu
+ \pmatrix { -1 \cr
	     \pm 3 \cr} b_{-{3\over 2}}^\mu \right]
|f^\mu =(1, \pm 1) \ket .}

For the minus sector, type I DGS is
\eqn\dgsbb{G^{-,I}_{2,\pm 1} \sim \left[
\pmatrix{ 3& \pm 2 \cr
	  \pm 2 & 1 \cr }
\alpha_{-1}^\mu b_{-\half}^\nu
+ \pmatrix { 1 \cr
	     \pm 1 \cr} b_{-{3\over 2}}^\mu \right]
|f^\mu =(-3, \pm 1) \ket ,}
and type II DGS is
\eqn\dgscc{G^{-,II}_{2,\pm 1} \sim \left[
\pmatrix{ 17& \pm 6 \cr
	  \pm 6 & 3 \cr }
\alpha_{-1}^\mu b_{-\half}^\nu
+ \pmatrix { 11 \cr
	     \pm 3 \cr} b_{-{3\over 2}}^\mu \right]
|f^\mu =(-3, \pm 1) \ket .}
Note that $3G^{-,I}_{2,\pm 1}- G^{-,II}_{2,\pm 1}$ is a ``pure $\bfi$'' DGS,
similar to the DGS in the bosonic Liouville theory.

We now apply the scheme used in  \refs{\cl} to derive
a general formula for the DGS.
{}From \winftm, the DGS in the minus sector can be written down explicitly
as follows
\eqn\gsmm{\eqalign{
\bg^{-}_{J,M} &\sim
\left[\int {d\bz} e^{- \bfi}(\bz)
\right] \bps^-_{J-1,M} (\boo)\cr
&\sim \bs_{2J-1}( -\bfi )
\bdel'(J-1,M,-i X) e^{[iM\bx + (-1-J)\bfi]}.}}
We thus have explicitly obtained a DGS for each $\bps^-$ discrete momentum.
However, there are still other DGS in this sector, for example,
the states
\eqn\gspphi{
\bg'^-_{J,M} \sim \left[\int d\bz
e^{-  \bfi (\bz)}\right]^{J-M} \bps^-_{M,M}(\boo)}
can be shown to satisfy the physical state
condtions. Since they are ``pure $\bfi$''
states, they are also DGS.
For example, $\bg_{1,0}^- =\bd\bfi e^{-\bfi}$
and $\bg_{2,\pm 1}^- =[- \bd^3\bfi +\bd\bfi\bd^2\bfi] e^{\pm i\bx-3\bfi}$,
which is a linear combination of \dgsbb\ and \dgscc .

\vskip 5mm

For the plus sector,
we can subtract two (distinct) positive norm discrete states at the
same momentum to obtain a pure gauge state
\eqn\gspps{
\bg^{+}_{J,M} =  (J+M+1)^{-1}
\int d\bz \left[\bps^+_{1,-1}(\bz) \bps^+_{J,M+1}(\boo) -
\bps^+_{J,M+1}(\bz) \bps^+_{1,-1}(\boo) \right].}
As an example, with \gspps\ one finds
\eqn\exam{\eqalign{
\bg ^+_{2, \pm 1} =& [\pm 3i\bd^3\bx +\bd^3\bfi +3i\bd^2\bx\bd\bx \cr
&\pm 2i\bd^2\bx\bd\bfi \pm 2i\bd\bx\bd^2\bfi+\bd\bfi\bd^2\bfi ]
e^{\pm i\bx + \bfi},}}
which is exactly the state we found in \dgsaa .
We thus have explicitly obtained a DGS for each $\bps^+$ momentum.

By construction in \gspps\ one can see that $\bg^+_{J,M}$ carry the
$w_\infty$ charges and serve as the symmetry parameters of the theory.
In fact, their operator products form the same $w_\infty$ algebra
\eqn\ginftp{ \int d\bz
\bg_{J_1,M_1}^+(\bz) \bg_{J_2,M_2}^+(\boo) = (J_2 M_1 - J_1 M_2)
\bg_{J_1+J_2-1,M_1+M_2}^+(\boo),}
where the RHS is defined up to another DGS.

\newsec {Conclusion}

We have demonstrated that the space-time $w_\infty$ symmetry parameters in
the $2D$ superstring theory come from solution of equations \gauone\
and \gautwo . This phenomenon should survive in the more realistic $10D$
heterotic string theory \refs{\lee} , although it would be difficult to
find the general solution (due to the high dimensionality of space-time).
The DGS in the old covariant quantization of the theory seems to be related
to the ground ring structure in the BRST approach. Finally,
the GSO projection can be easily imposed on the spectrum.

\vskip 1cm

\noindent
{\bf Acknowledgements.}

This research is financially supported by National Science Council of Taiwan,
R.O.C., under grant number NSC84-2811-M009-006 and NSC83-0208-M-009-063.

\vskip 1cm
\vfill\supereject
\listrefs

\end